\documentclass[%
 reprint,
 amsmath,amssymb,
 aps,
]{revtex4-2}

\usepackage{graphicx}
\usepackage{dcolumn}
\usepackage{bm}

\usepackage[version=3]{mhchem}
\usepackage{booktabs}
\usepackage{multirow}

\usepackage{upgreek}
\usepackage{xcolor}
\definecolor{darkblue}{RGB}{25,88,128}
\usepackage[
    colorlinks=true,
    linkcolor=darkblue,
    filecolor=darkblue,
    citecolor=darkblue,      
    urlcolor=darkblue,
    ]{hyperref}

\begin{document}

\preprint{APS/123-QED}

\title{Interplay between Ferroelectricity and Metallicity in \ce{BaTiO3}}

\author{Veronica F. Michel}
\author{Tobias Esswein}\email{tobias.esswein@mat.ethz.ch}
\author{Nicola A. Spaldin}\email{nicola.spaldin@mat.ethz.ch}
\affiliation{%
 Materials Theory, Department of Materials, ETH Zürich, Wolfgang-Pauli-Strasse 27, 8093 Zürich, Switzerland
}%

\date{\today}

\begin{abstract}
We explore the interplay between ferroelectricity and metallicity, which are generally considered to be contra-indicated properties, in the prototypical ferroelectric barium titanate, \ce{BaTiO3}. Using first-principles density functional theory, we calculate the effects of electron and hole doping, first by introducing a hypothetical  background charge, and second through the introduction of explicit impurities (La, Nb and V for electron doping, and K, Al and Sc for hole doping). We find that, apart from a surprising increase in polarization at small hole concentrations, both charge-carrier types decrease the tendency towards ferroelectricity, with the strength of the polarization suppression, which is different for electrons and holes, determined by the detailed structure of the conduction and valence bands. Doping with impurity atoms increases the complexity and allows us to identify three factors that influence the ferroelectricity: structural effects arising largely from the size of the impurity ion, electronic effects from the introduction of charge carriers, and changes in unit-cell volume and shape. A competing balance between these contributions can result in an increase or decrease in ferroelectricity with doping.
\end{abstract}

\maketitle


\section{Introduction}
Ferroelectricity and metallicity are generally considered to be contra-indicated properties since the metallic charge carriers screen the long-range interactions that favor the ferroelectric structural distortion~\cite{Book_Lines}. These two properties can nevertheless be combined in systems where the interaction between the itinerant electrons and the polar distortion is weak, as shown theoretically by Anderson and Blount in 1965. Such materials were called ‘ferroelectric metals’~\cite{Anderson1965}, although the term is somewhat ambiguous: Ferroelectricity is in fact defined for materials that show a spontaneous polarization that is \textit{switchable by an applied electric field}; in the presence of metallic charge carriers however, the electric field is screened by the itinerant electrons and induces an electric current rather than a polarization switch~\cite{Cordero2019,Wang2012}, so that a ferroelectric can formally not be metallic. A more rigorous term is perhaps polar metal, which is a material, such as \ce{LiOsO3}~\cite{Shi2013}, that has a polar crystal class combined with a non-zero density of states at the Fermi level~\cite{Benedek2016}. In addition to their fundamental interest, polar metals are promising for a range of applications. Some of them are reported to show giant optical responses, and could thus be used in optoelectronic devices~\cite{Kim2016}. They are also good candidates for the design of materials with tunable metal-insulator transitions~\cite{Puggioni2015}. Furthermore, they are of relevance for polar superconductors, which have the potential to show unconventional superconducting states~\cite{Smidman2017}.

Barium titanate (\ce{BaTiO3}) is the prototypical ferroelectric material. It has the ideal \ce{ABO3} perovskite structure at high temperatures, corresponding to a centrosymmetric cube with formally Ba$^{2+}$ cations at the corners, Ti$^{4+}$ at the center and O$^{2-}$ at the face centers~\cite{Kwei1993}. Upon lowering the temperature, three phase transitions occur: At 393~K a phase transition from a paraelectric cubic to a ferroelectric tetragonal phase is observed, followed by a transition to a ferroelectric orthorhombic phase at 278~K and to a ferroelectric rhombohedral phase, with polarization along a $<$111$>$ direction, at 183~K~\cite{Kwei1993}. Multiple efforts have been made to make \ce{BaTiO3} metallic. In 2008, an insulator-metal transition was reported in \ce{BaTiO3} doped with oxygen vacancies (BaTiO$_{3-\delta}$)~\cite{Kolodiazhnyi2008}, which was later shown to retain its ferroelectric structural distortion up to an electron concentration n~$\approx$~1.9$\cdot$10$^{21}$~cm$^{-3}$~\cite{Kolodiazhnyi2010}. This work was challenged by Jeong~\textit{et~al.}, who reported, based on neutron diffraction studies, that the ferroelectric ordering and the metallic conduction are not coexisting but rather form two distinct phases in BaTiO$_{3-\delta}$~\cite{Jeong2011}. Recently, Cordero~\textit{et~al.} showed through elastic response studies that the ferroelectric transitions persist in metallic BaTiO$_{3-\delta}$~\cite{Cordero2019}, confirming the observations of Ref.~\cite{Kolodiazhnyi2010}. In addition to the introduction of oxygen vacancies, doping in \ce{BaTiO3} can be achieved through atomic substitutions, the most common being La$^{3+}$ for Ba$^{2+}$ and Nb$^{5+}$ for Ti$^{4+}$. Both La- and Nb-doped \ce{BaTiO3} show electrical conductivity with a polaronic conduction mechanism~\cite{Iguchi1991,Gillot1992}, with the polaron formation in Nb-doped \ce{BaTiO3} believed to come from incoherent B-site off-centering resulting from the random Nb substitution~\cite{Page2008}.

A few computational investigations on doped \ce{BaTiO3} have been reported in the last decade. In 2012, Wang~\textit{et~al.} studied metallic \ce{BaTiO3}, introducing charge carriers through the background-charge method described below, and showed that the ferroelectric displacements are sustained up to 0.11 electrons per unit cell (0.11~e/u.c.)~\cite{Wang2012}. In the same year, Iwazaki~\textit{et~al.} also investigated background-charge doped \ce{BaTiO3}, as well as Nb substitution and oxygen vacancies as donors~\cite{Iwazaki2012}. Note that the calculations in both these works were performed with fully relaxed lattice constants, that were shown to be ill-defined with the background-charge method by Bruneval~\textit{et~al.} in 2015~\cite{Bruneval2015}. In 2016, Benedek~\textit{et~al.} compared the effect of doping in \ce{LiOsO3} and \ce{ATiO3} perovskites (A = Ba, Sr, Ca)~\cite{Benedek2016}. They showed that the ferroelectricity is suppressed by electron doping in \ce{BaTiO3}, whereas the non-centrosymmetricity in metallic \ce{LiOsO3} and pseudocubic \ce{CaTiO3} (with octahedral rotations not allowed) is robust to addition of charge carriers because of the local-bonding nature of the mechanism underlying the off-centering of the ions. Recently, a meta-screening effect was proposed by Zhao~\textit{et~al.} as the main factor determining the persistence of the polar phase in metallic ferroelectrics~\cite{Zhao2018}.

In this work, we explore theoretically the interplay between the contra-indicated properties of ferroelectricity and metallicity in \ce{BaTiO3}. Using first-principles density functional theory, we address the question of how the ferroelectric B-site off-centering and the likelihood of polarization switchability are affected by metallic charge carriers in \ce{BaTiO3}. We investigate both electron and hole doping in \ce{BaTiO3} first through the background-charge method and subsequently by introducing explicit dopants (La, Nb, V, K, Al and Sc). The dopants are chosen based on experimental feasibility and allow us to separate the effects of dopant size, substitution site as well as second-order Jahn-Teller (SOJT) activity on the polarization.

\section{Computational Method}

Our calculations are performed using density functional theory (DFT) as implemented in the VASP code~\cite{Kresse1999}, using the recommended projector augmented wave potentials and the PBEsol exchange-correlation functional~\cite{Perdew2008}. We include plane waves up to a kinetic energy cutoff of 600~eV. All the used k-point grids are Gamma-centered; details on the sampling meshes are given below. We converge total free energies to 10$^{-6}$~eV and relax atomic positions until all force components converge below 10$^{-3}$~eV/Å.

Crystal structure relaxations are computed for both single unit cells and supercells of \ce{BaTiO3}. We consider both tetragonal and cubic \ce{BaTiO3} single unit cells, corresponding to the room-temperature and high-temperature structures, respectively. We compute the tetragonal \ce{BaTiO3} single unit cell structural relaxation as follows: We start with the experimental high-temperature, cubic structure and fully relax its atomic positions and lattice constants with a 6$\times$6$\times$6 k-points grid, while retaining the cubic symmetry. We subsequently displace the Ti atom by 1$\%$ along the $c$ direction and fully relax the atomic positions and the lattice constants with a 12$\times$12$\times$12 k-point grid, while retaining the tetragonal symmetry. We obtain $a$~=~$b$~=~3.967~Å and $c$~=~4.065~Å; these lattice constants are consistent with experimental literature values for the room-temperature tetragonal phase~\cite{Buttner1992}. We apply the same procedure for the cubic \ce{BaTiO3} single unit cell but fix the lattice constants for the second structural relaxation. This keeps the cubic equal lattice constants while allowing for a lower internal symmetry to permit a ferroelectric distortion; we refer to this unit cell as pseudocubic throughout the manuscript. We construct the supercells as multiples of the single unit cell in the $a$, $b$ and $c$ directions and we investigate 2$\times$2$\times$2 supercells in detail, for which we use a 12$\times$12$\times$12 k-point grid. 

We compute densities of states for \ce{BaTiO3} single unit cells with a 24$\times$24$\times$24 k-point grid. Crystal Orbital Hamiltonian Populations are calculated using the LOBSTER package~\cite{Dronskowski1993,Maintz2013,Deringer2011,Maintz2016,Nelson2020,Maintz2016a}. Our basis sets for COHP calculations are Ba (5s, 5p, 5d, 6s), Ti (3s, 3p, 3d, 4s) and O (2s, 2p). A 16$\times$16$\times$16 k-point grid is used for those calculations.

We dope the \ce{BaTiO3} single unit cells with the background-charge method, in which the number of electrons in the system is manually adapted to the desired doping level and a uniform background charge is added to enforce charge neutrality~\cite{Guide_vasp}. Note that the relaxation of lattice constants is not well-defined when using the background-charge method~\cite{Bruneval2015}, so we keep them fixed. We dope supercells by introducing impurity atoms and allowing for relaxation of the lattice constants.

The presented ‘polarization’ values are calculated using

\begin{equation}
    \delta P = \frac{e}{\Omega}\sum_iZ^*_id_i,
\end{equation}

\noindent
where $e$ is the electronic charge, $\Omega$ the unit cell volume, $Z^*_i$ the Born effective charge (BEC) of atom $i$, and $d_i$ its relative displacement in the ferroelectric direction~\cite{Spaldin2012}. The BECs we use are +2.7 for barium, +7.25 for titanium, -5.71 for oxygens that displace parallel to their Ti-O bond direction and -2.15 for oxygens that displace perpendicular to this direction~\cite{Ghosez1995}. This procedure yields a polarization value of 38~$\upmu$C/cm$^2$ for undoped \ce{BaTiO3}, which is close to the value of 34~$\upmu$C/cm$^2$ that we compute using the Berry-phase method with a 24$\times$24$\times$24 k-point grid (see Figure \ref{fig:berry_phase} in the SI), in agreement with literature values~\cite{Yuk2017,Wahl2008a}. For impurity-doped supercells we set the BEC of the impurity to that of the atom it substitutes. Note that the polarization of the doped (metallic) systems is ill-defined as it cannot be switched. This quantity does not correspond to a true polarization obtained through a Berry-phase approach~\cite{King-Smith1993,Resta1993}, is not necessarily switchable and does not require insulating behavior. Its value reflects, rather, the amount of ferroelectric-like structural distortion present in the system. Throughout this paper, we use the term polarization to refer to this effective polarization, and the term ferroelectricity to refer to the non-centrosymmetric structural distortion, without implying polarization switchability.

\section{Results and Discussion}

\subsection{Introduction of Background Charge into a \texorpdfstring{\ce{BaTiO3}}{BaTiO3} Unit Cell}

In the first part of our study, we investigate the introduction of doping in \ce{BaTiO3} through the background-charge method. This approach allows us to isolate the effect of the electronic charge from other influences such as ion size or change in chemistry associated with the introduction of explicit dopant atoms. We find the ferroelectricity to be reduced by the introduction of charge carriers, and to be more sensitive to the introduction of electrons than holes, due to the different character of the valence and conduction bands.

\begin{figure}
    \centering
    \includegraphics[width=0.9\linewidth, keepaspectratio]{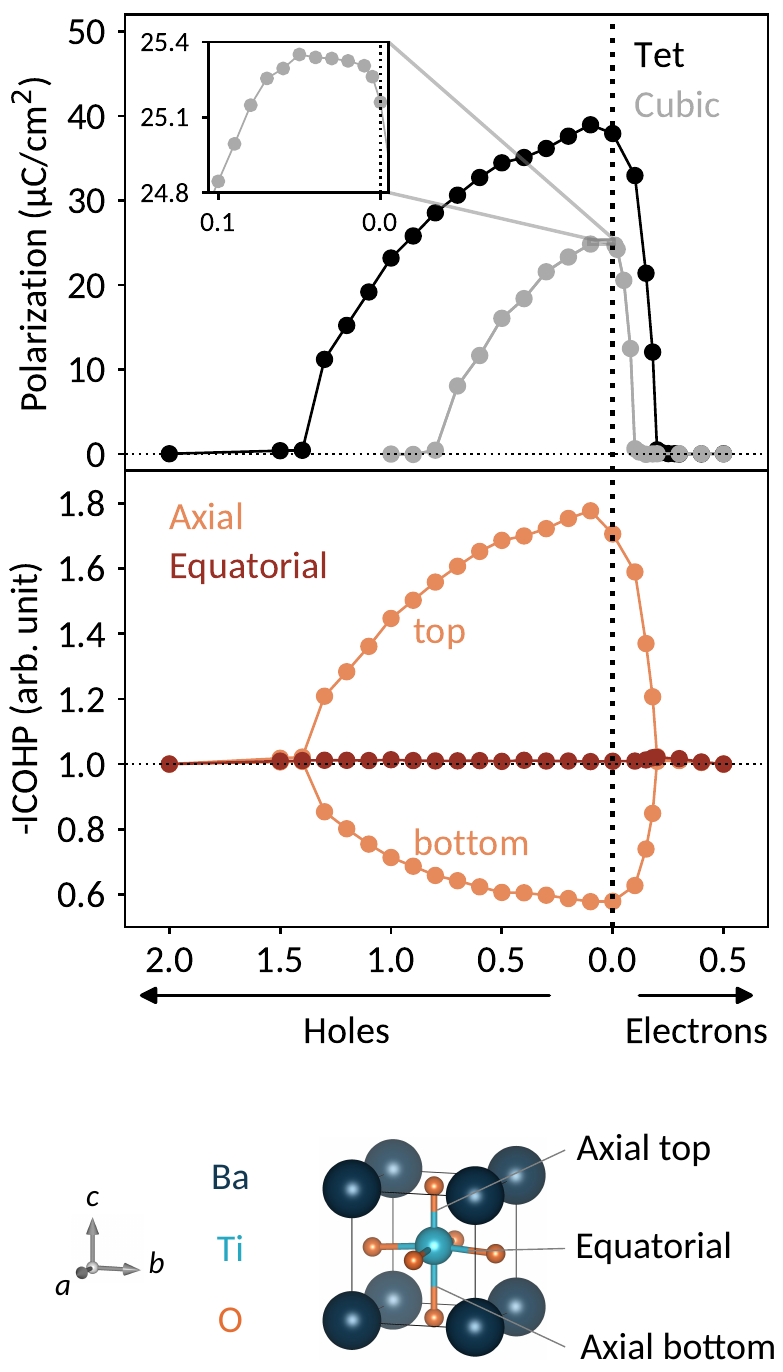}
    \caption{Top: Polarization as a function of the charge-carrier concentration (electrons on the right, holes on the left) for tetragonal and pseudocubic (abbreviated \textit{tet} and \textit{cubic}) \ce{BaTiO3}. The pseudocubic symmetry refers to equal lattice constants with an allowed lower internal symmetry so that a ferroelectric distortion can occur. We see that the polarization is reduced on electron doping for all concentrations. The addition of holes first increases the polarization before reducing it, with a slower polarization suppression than in the electron doping case. The pseudocubic and tetragonal systems behave in a very similar manner, although the pseudocubic system has lower polarization values for the same charge-carrier concentration. Bottom: Integrated Crystal Orbital Hamiltonian Populations (ICOHPs) for the Ti-O bonds of tetragonal \ce{BaTiO3} as a function of the charge-carrier concentration. Three Ti-O bonds are considered and showed in the figure: the Ti-O axial top, axial bottom and equatorial. All ICOHPs are normalized to their value in the centrosymmetric structure. The higher the ICOHP value, the stronger its corresponding bond.}
    \label{fig:polarization_icohp_bg}
\end{figure}

\subsubsection{Electron Doping}

The calculated polarization of \ce{BaTiO3} as a function of charge-carrier concentration is shown in the top panel of Figure~\ref{fig:polarization_icohp_bg}, for both electron and hole doping, and tetragonal and cubic unit cells. We discuss the electron-doped tetragonal system first. On electron addition, the polarization of tetragonal \ce{BaTiO3}  decreases until it is completely suppressed at 0.2~e/u.c.

\begin{figure}
    \centering
    \includegraphics[width=0.85\linewidth, keepaspectratio]{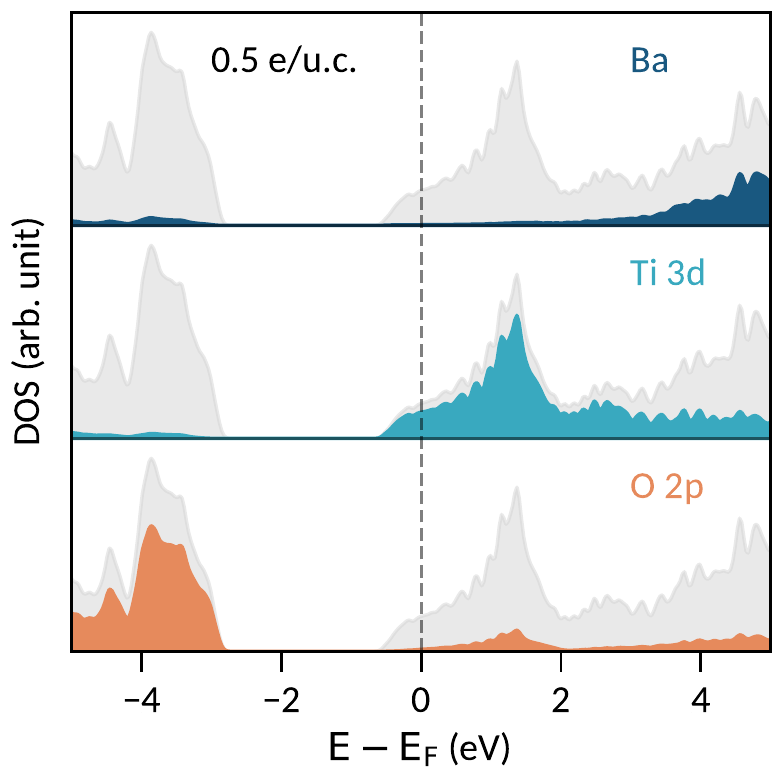}
    \caption{Density of states for tetragonal \ce{BaTiO3} doped with 0.5~e/u.c. The total DOS is shown in gray and the Ba, Ti 3d and O 2p contributions are shown in blue, teal and orange, respectively. The valence band has mainly O 2p character, whereas the conduction band has a high Ti 3d contribution. The Fermi energy, E$_\mathrm{F}$, lies close to the bottom of the conduction band due to the addition of electrons.}
    \label{fig:dos_tet_bg_0.5e}
\end{figure}

The decrease in ferroelectricity can be understood through simple electronic structure considerations: In Figure~\ref{fig:dos_tet_bg_0.5e} we show the calculated density of states in the region of the Fermi energy for tetragonal \ce{BaTiO3} doped with 0.5~e/u.c. We see that the valence band has mainly O~2p character, whereas the conduction band has mainly Ti~3d character and the barium ions have no significant contribution to the DOS in the region of the band gap. Compared to pure \ce{BaTiO3} (not shown), the DOS is not strongly affected by the charge carriers, apart from the shift of the Fermi level into the conduction band. 

\begin{figure*}
    \centering
    \includegraphics[width=0.9\linewidth, keepaspectratio]{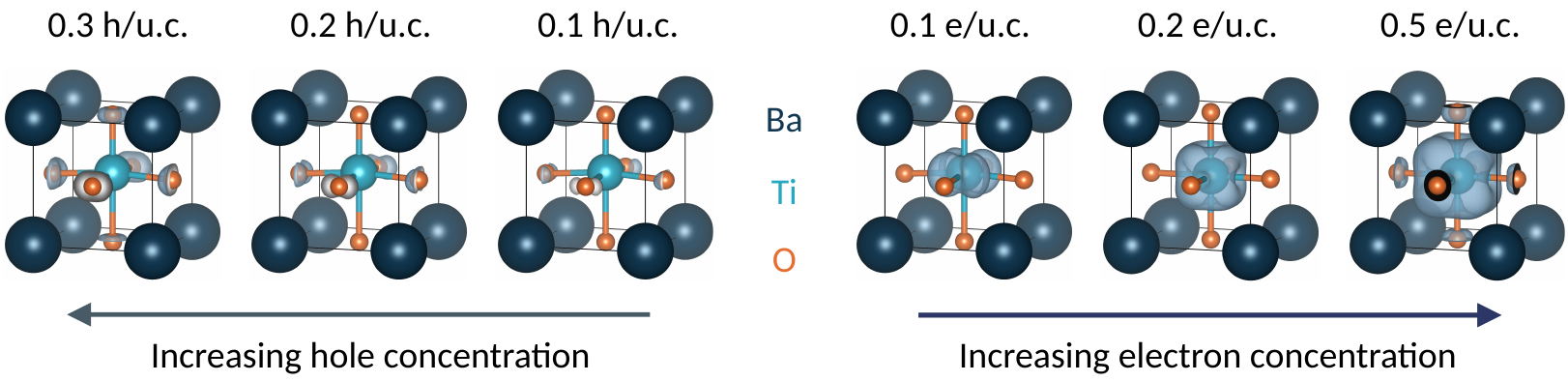}
    \caption{Conduction-band electron charge density (right) and valence-band hole charge density (left) for tetragonal \ce{BaTiO3}. For electron doping, the gray surfaces all have the same isosurface level (charge~$\approx$~7.5~$\cdot$~10$^{-4}$~\textit{e}) and show the location of the added electrons. For hole doping, the gray surfaces also have the same isosurface level, which is one order of magnitude larger than for electrons (charge~$\approx$~7.3~$\cdot$~10$^{-3}$~\textit{e}), and reveal where the electron depletion occurs.}
    \label{fig:band_decomposed_charge_densities_bg}
\end{figure*}

The added electrons occupying primarily the Ti~3d energy levels can also be seen clearly in the charge densities of Figure~\ref{fig:band_decomposed_charge_densities_bg}, where the gray surfaces indicate the added electron density. With increasing electron doping, we see that the charge density on the Ti 3d orbitals progressively increases and as a result the Ti ion deviates further from its formally d$^0$ electron configuration. Correspondingly, the off-centering described by the SOJT effect, which is favored for d$^0$ electronic configurations, is reduced~\cite{Rondinelli2009}.

We compute integrated Crystal Orbital Hamiltonian Populations (ICOHPs) to confirm our explanation for the ferroelectricity decrease with increasing electron concentration. The bottom of Figure~\ref{fig:polarization_icohp_bg} shows the ICOHP as a function of the charge-carrier concentration for the Ti-O axial top, axial bottom and equatorial bonds (these are also illustrated in Figure~\ref{fig:polarization_icohp_bg}). All the values are normalized to their respective value in the centrosymmetric structure. In the ferroelectric distorted structure, the Ti is off-centered along $c$, resulting in a strong, short Ti-O axial top bond with a high ICOHP value and a weak, long bottom bond with a low ICOHP value. Upon adding electrons, the top bond is weakened and its ICOHP value is lowered. Conversely, the bottom bond is strengthened and its ICOHP value increases. The Ti-O equatorial bonds are not particularly affected by the structural distortion decrease.

\subsubsection{Hole Doping}
In contrast to electron doping, the introduction of holes in tetragonal \ce{BaTiO3} causes first an increase in ferroelectricity (from 0 to 0.2~h/u.c., as seen in the top of Figure~\ref{fig:polarization_icohp_bg}), followed by a decrease ($>$ 0.2 h/u.c.) and complete suppression at 1.5~h/u.c. Overall, the amount of charge carriers needed to suppress the ferroelectric distortion is a factor 7.5 larger for hole doping than for electron doping.

\begin{figure}[h]
    \centering
    \includegraphics[width=0.85\linewidth, keepaspectratio]{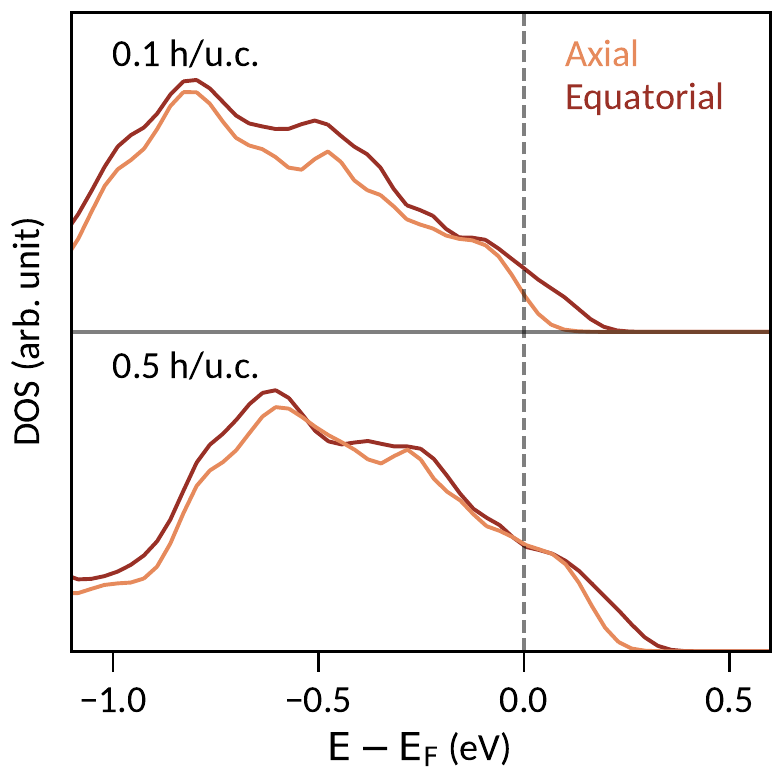}
    \caption{Oxygen site-resolved densities of states for tetragonal \ce{BaTiO3} doped with 0.1~h/u.c. (top) and 0.5~h/u.c. (bottom). At 0.1~h/u.c., electrons are primarily removed from the equatorial oxygen atoms. At 0.5~h/u.c., the axial oxygen atoms are also strongly depleted.}
    \label{fig:dos_tet_bg_h}
\end{figure}

The particular behavior of the hole-doped systems can be understood by considering their detailed densities of states. Figure~\ref{fig:dos_tet_bg_h} shows the oxygen site-resolved DOS for 0.1 and 0.5~h/u.c., with axial and equatorial sites shown separately. On removing electrons, the Fermi energy shifts down into the valence band, corresponding to a charge depletion of the oxygen atoms. For doping concentrations less than 0.1~h/u.c., we see (Figure \ref{fig:dos_tet_bg_h}, top panel) that electrons are largely only removed from the equatorial oxygen band, and the axial oxygens are almost unaffected. This electron depletion weakens the equatorial Ti-O bonds, and reduces the tension in the equatorial plane allowing the ferroelectric displacements along $c$ to increase. For higher charge-carrier concentrations, depletion from the axial oxygen atoms starts to take place. We show in Figure~\ref{fig:dos_tet_bg_h} the case of 0.5~h/u.c. and see that the Fermi energy lies in both the axial and equatorial bands. Again invoking the SOJT effect, in which axial oxygens transfer electrons to the empty Ti d orbitals and stabilize the off-centering, we see that removing electrons from the axial oxygens reduces the tendency for the ferroelectric structural distortion to occur. The evolution of the electron depletion from equatorial to axial oxygens with increasing hole concentration can also be seen in the charge densities of Figure~\ref{fig:band_decomposed_charge_densities_bg}.

The effect of the electron depletion on the Ti-O bond is also seen in the ICOHP which is plotted as a function of the hole concentration in Figure~\ref{fig:polarization_icohp_bg}. Between 0 and 0.2~h/u.c., the Ti-O axial bottom bond is weakened and the Ti-O axial top bond is strengthened, as seen in their respective decreasing and increasing ICOHP values, supporting the explanation for the increase in ferroelectricity proposed above. For 0.3~h/u.c. and larger, a behavior similar to the electron doping case is observed, consistent with the observed loss of ferroelectricity. The ICOHP values for the Ti-O axial top bond follow the trend of the polarization values, indicating a direct relation between changes in polarization and Ti-O bond strength.

\subsubsection{Influence of Unit Cell Shape}

Next we investigate the influence of the unit cell shape by comparing the behavior of tetragonal and pseudocubic (abbreviated \textit{tet} and \textit{cubic} in Figure~\ref{fig:polarization_icohp_bg}) \ce{BaTiO3}. Note that, as mentioned above, the pseudocubic system has equal lattice parameters and therefore cubic shape, but we allow for a lower internal symmetry to permit a ferroelectric distortion. The tetragonal and pseudocubic systems follow the same trends, as seen in the top of Figure~\ref{fig:polarization_icohp_bg}. Their similarity can be explained by their comparable electronic structures (the DOS for the pseudocubic system can be found in Figures \ref{fig:dos_cub_bg_0.5e} and \ref{fig:dos_cub_bg_h} of the SI). The polarization of the pseudocubic system is, however, smaller than the tetragonal one and its polarization is suppressed faster. This is related to the absence of coupling to strain, and correspondingly smaller space in the $c$ direction, resulting in smaller atomic displacements and thus a smaller polarization. In the rest of the paper, we present only the results for tetragonal \ce{BaTiO3}. Note that we focus on the tetragonal symmetry to provide guidelines on the behaviour of \ce{BaTiO3} at room temperature, and that we do not expect the mechanisms to be particularly different for other symmetries, for example the rhombohedral that occurs at low temperatures.

\subsection{Introduction of Impurity Atoms in \texorpdfstring{\ce{BaTiO3}}{BaTiO3} Supercells}

In the second part of our study, we include explicit impurity atoms in tetragonal \ce{BaTiO3} and investigate their effect on the polarization of the system. We consider electron and hole doping through A-site as well as B-site substitution. For electron doping, we introduce La on the A~site and Nb or V on the B~site. We explore hole doping through the introduction of K on the A~site and Al or Sc on the B~site. All dopants of interest add or remove one charge carrier per 2$\times$2$\times$2 supercell, resulting in carrier concentrations of 0.125~carriers/u.c. Results for smaller charge-carrier concentrations can be found in Figure \ref{fig:pol_vol_c_a_supercells} of the SI. Note that we consider only single substitutional impurities that introduce either one electron or one hole in their usual formal charge state. We do not investigate vacancies, the introduction of multiple defects or impurity-vacancy complexes.

\begin{figure*}
    \centering
    \includegraphics[width=0.9\linewidth, keepaspectratio]{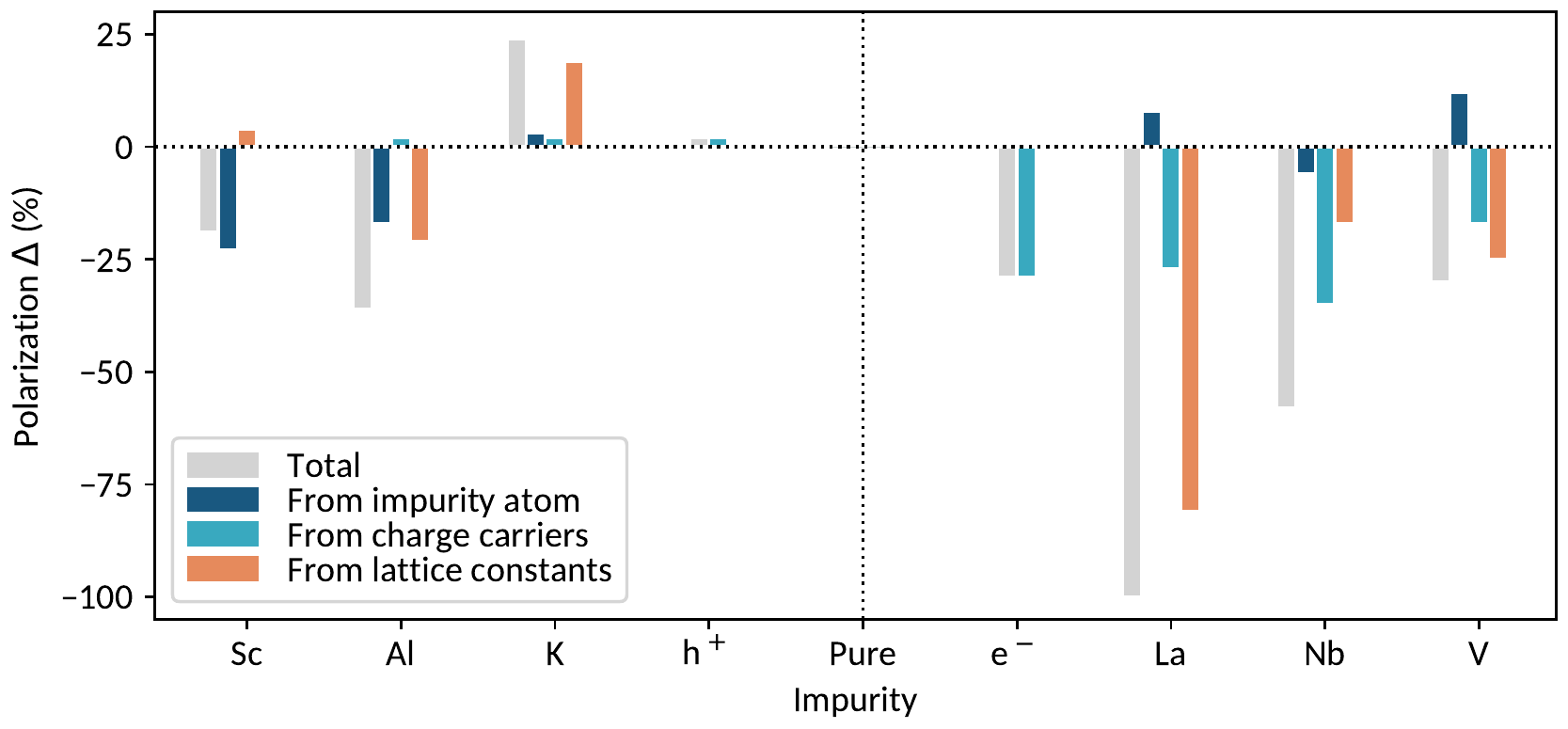}
    \caption{Polarization difference caused by the introduction of doping in tetragonal \ce{BaTiO3}. All the systems have a charge-carrier concentration of 0.125~carriers/u.c., corresponding to one impurity atom in a 2$\times$2$\times$2 supercell. Positive polarization $\Delta$ values (y axis) correspond to a polarization gain, negative ones to a polarization loss. The total polarization difference (gray) is split into contributions corresponding to: i) The effect of the impurity atom without its carriers (dark blue). ii) The effect of the introduced charge carriers (teal). iii) The effect of the changing shape and volume of the system (orange). The total polarization change is obtained as the sum of the competing contributions present in the system.}
    \label{fig:polarization_change_percentage}
\end{figure*}

For every X-\ce{BaTiO3} system (where X denotes a general dopant), we perform three different calculations in order to identify and separate contributions to the polarization change. The first calculation consists of a structural relaxation of the internal coordinates of X-\ce{BaTiO3} containing the impurity atom but not its corresponding charge carriers and with lattice constants fixed to their undoped values. We achieve this by compensating the charge carriers added with the dopant by background-charge doping (the background charge requires the use of fixed lattice constants). We refer to this component as the \textit{Impurity-atom contribution}. In the second calculation we remove the compensating background-charge and relax the internal coordinates of X-\ce{BaTiO3} with the impurity atom and its charge carriers, still keeping the lattice constants fixed. We refer to this as the \textit{Charge-carrier contribution}. In the third calculation, we make a full structural relaxation of both the internal coordinates and lattice parameters of X-\ce{BaTiO3}, containing the impurity atom with its charge carriers; we call this situation \textit{free lattice constants} and we refer to this third component as the \textit{Lattice-constant contribution}. This last system corresponds to the most realistic one. These three scenarios allow us to separate the contributions to the polarization change coming from the impurity atom, from the charge carriers and from changes in lattice constants. In the next sections, we present a detailed analysis of these contributions for the different impurity-doped \ce{BaTiO3} systems. A compact overview of the discussed systems and features can be found in Table~\ref{tab:overview}. The polarization changes arising from the different contributions are illustrated in Figure~\ref{fig:polarization_change_percentage}. Background-charge results are given as a reference in both cases and all given polarization changes $\Delta$P in $\%$ refer to the changes compared to the original \ce{BaTiO3} polarization.

Note that all the listed Shannon ionic radii are taken from Refs.~\cite{Shannon1976, Site_shannon}. and that tolerance factors are calculated with these values, with the charge of the impurity ion taken as its formal charge when introduced as a dopant -- e.g. 5+ for Nb -- with the coordination number appropriate to the substituted site (12 for the A~site and 6 for the B~site).

\begin{table*}
\centering
\caption{Overview of the investigated doped \ce{BaTiO3} systems and their respective calculated total polarization changes $\Delta$P. The dopants are listed with specification of their type and site. \textit{Background} refers to doping through the background-charge method and is intended to serve as a reference. The contributions from the impurity atom, the charge carriers and the free lattice constants are highlighted. The size of the impurity atom refers to its Shannon ionic radius. The reference ionic radii of Ba$^{2+}$, Ti$^{4+}$ and O$^{2-}$ are 1.61~Å, 0.61~Å and 1.35~Å, respectively. The systems can be described as \textit{non-polar} or \textit{polar}}
\label{tab:overview}
\begin{tabular*}{\textwidth}{@{\extracolsep{\fill}}lccccccccccc@{}}
    \toprule
    \multirow{2}{*}{Doping type} & \multirow{2}{*}{Dopant} & \multirow{2}{*}{Site} &\multicolumn{2}{c}{Impurity atom} & \multicolumn{2}{c}{Charge carriers} & \multicolumn{2}{c}{Lattice constants} & \multirow{2}{*}{Total $\Delta$P ($\%$)} & \multirow{2}{*}{Description}\\
    \cmidrule(r){4-5}
    \cmidrule(r){6-7}
    \cmidrule(r){8-9}
    & & & Size (Å) & SOJT & Type & Localization & Volume & Tetragonality & & \\
    \midrule
    \multirow{5}{*}{Donors} & Background & - & - & - & e$^-$ & no & - & - & -- 29 & - \\
    & La$^{3+}$ & A & 1.36 & - & e$^-$ & no & $\downarrow$ & loss & -- 100 & non-polar \\
    & Nb$^{5+}$ & B & 0.64 & yes & e$^-$ & yes & const. & loss & -- 58 & polar \\
    & V$^{5+}$ & B & 0.54 & yes & e$^-$ & yes & $\downarrow$ & loss & -- 30 & polar \\
    \midrule
    \multirow{4}{*}{Acceptors} & Background & - & - & - & h$^+$ & no & - & - & + 2 & - \\
    & K$^{+}$ & A & 1.64 & - & h$^+$ & no & const. & gain & + 24 & polar \\
    & Al$^{3+}$ & B & 0.54 & no & h$^+$ & no & $\downarrow$ & loss & -- 36 & polar \\
    & Sc$^{3+}$ & B & 0.75 & yes & h$^+$ & no & $\uparrow$ & loss & -- 19 & polar \\
    \bottomrule
\end{tabular*}
\end{table*}

\subsubsection{Donor Doping}
\paragraph{A-Site Dopants}
We start by considering the La-doped \ce{BaTiO3} system (Ba$_{1-x}$La$_{x}$TiO$_3$, BLTO). The La substitutes on the A~site as formally trivalent La$^{3+}$, which has a smaller ionic radius than Ba$^{2+}$ and, as we discuss below, influences therefore the structure of the system.
\\
\textit{Impurity-atom contribution} When a La atom without its respective charge carriers is added to \ce{BaTiO3}, we find that the polarization of the system increases by 8~$\%$, due to a displacement of the small La$^{3+}$ cation in the $c$ direction. Besides directly contributing to the polarization of the system, the La also has an influence on the Ti~sites in two ways: First, the La is underbound and pulls the Ti atoms immediately above it towards itself, reducing their off-centering. Second, the La displacement along $c$ leaves more space for the Ti below it to off-center; this contribution dominates over the first.
\\
\textit{Charge-carrier contribution} Upon adding the electrons, the polarization of the system is reduced by 27~$\%$ compared to the original \ce{BaTiO3} polarization. This value is close to that found in the background-charge doped reference (29~$\%$). In fact, the introduction of the La does not influence the density of states of the system. The La f states lie high in energy and do not affect the region around the Fermi energy; the bottom of the conduction band therefore has pure Ti character and the electrons distribute homogeneously over the B~sites of the system.
\\
\textit{Lattice-constant contribution} Finally we consider the relaxation of the lattice constants. Here we find that the decreases in both volume and tetragonality, due to the small La size, reduce the polarization by a further 81~$\%$.
\\
\textit{Summary} The introduction of La in \ce{BaTiO3} results in a complete polarization suppression with all the atoms adopting their centrosymmetric positions. The volume and tetragonality decrease caused by the small La size are the main factors dominating the polarization loss. Experimental reports claim that the conduction mechanism in BLTO is of polaronic nature~\cite{Iguchi1991}. No polaronic localization is seen in the above presented results. This could be due to an incomplete description of the charge imbalance compensation mechanisms present in the system. We consider only an electronic compensation of Ba$^{2+}$ $\Rightarrow$ La$^{3+}$ + e$^-$, without any further ionic contributions as reported in the literature (e.g. 4Ba$^{2+}$ + Ti$^{4+}$ $\Rightarrow$ 4La$^{3+}$ + V$_\mathrm{Ti}$)~\cite{Morrison2001}.

\paragraph{B-Site Dopants}

We consider Nb and V as B-site substituting donor impurities; the respective systems can be written as BaTi$_{1-x}$Nb$_x$O$_3$ (BTNO) and BaTi$_{1-x}$V$_x$O$_3$ (BTVO). When introduced in \ce{BaTiO3}, Nb and V are present as Nb$^{5+}$ and V$^{5+}$, respectively, and have therefore formally empty d orbitals, making them SOJT active. Nb$^{5+}$ (0.64~Å) is larger than Ti$^{4+}$ (0.61~Å), whereas V$^{5+}$ (0.54~Å) is smaller. The off-centering of the two dopants depends on their size and the tolerance factor of the structure.

We begin with the Nb-doped \ce{BaTiO3} system:
\\
\textit{Impurity-atom contribution} The introduction of the Nb without its charge carriers decreases the polarization by 6~$\%$, due to the reduced off-centering of the Nb atom and its axial neighboring Ti along the polar axis. The reduced structural distortion is due to the Nb size: The tolerance factor of \ce{BaNbO3} is 1.03, compared to 1.07 for \ce{BaTiO3}. Although it has the empty d orbital configuration needed to be SOJT active, it is too big to actually off-center. The Ti atoms axial to the Nb are directly affected by the Nb behavior: Since the Nb is less off-centered, they have less space to displace in the $c$ direction and are more centrosymmetric.
\\
\textit{Charge-carrier contribution} The addition of the electrons reduces the polarization by a further 35~$\%$, which is more than expected from the background-charge reference (29~$\%$). The additional suppression results from a slight accumulation of the electrons on the Nb atom and its axial neighboring Ti along the polar axis, which have the largest contribution to the bottom of the conduction band (Figure \ref{fig:bond_ratios_supercells} of the SI). The electron accumulation heavily reduces the structural distortion of these two sites, increasing the polarization loss due to the presence of the charge carriers.
\\
\textit{Lattice-constant contribution} When the lattice constants are relaxed, the polarization is reduced by a further 17~$\%$. This polarization change can be assigned to the reduced tetragonality since the volume of the system stays rather constant. Note that this constant volume is not directly intuitive. In fact, Nb$^{5+}$ is larger than Ti$^{4+}$, so that a structural expansion would be expected. It is nevertheless counterbalanced by the loss of ferroelectricity, resulting in an unchanged volume.
\\
\textit{Summary} Nb-doped \ce{BaTiO3} remains polar, although with an overall polarization decrease of 58~$\%$. The charge carriers are the most important factor contributing to the polarization loss. We find that they tend to accumulate on the Nb atom and its nearest axial Ti, consistent with the polaronic conduction mechanism proposed in the literature~\cite{Gillot1992,Page2008}.

Next we evaluate the polarization change in the V-doped system where we find that the small size of V leads to markedly different behavior.
\\
\textit{Impurity-atom contribution} The addition of V without its respective charge carriers induces a polarization increase of 12~$\%$: The V is smaller than the Ti and therefore able to off-center more (\ce{BaVO3} has a large tolerance factor of 1.11). The Ti axial to the V are also off-centered more because the V influences its axial environment and induces a cascade off-centering -- the same effect but in the reverse direction as that seen for the BTNO system.
\\
\textit{Charge-carrier contribution} The introduction of electrons decreases the polarization by 17~$\%$ compared to pure \ce{BaTiO3}, which is less than expected from the background-charge reference (29~$\%$). The added charge carriers strongly localize on the V because of its very large contribution to the bottom of the conduction band (Figure \ref{fig:dos_supercells} of the SI). We might expect this localization to increase the polarization suppression, as seen in the BTNO system. However, in this case the V is so small that it displaces and contributes to the polarization even if its SOJT hybridization stabilization is quenched by the extra electrons. In addition, the electron localization on the V reduces their accumulation on the Ti~sites, so that their polar structural distortions are not affected much by the doping.
\\
\textit{Lattice-constant contribution} The relaxation of the lattice constants results in a further polarization decrease of 25~$\%$. This is explained by the reduced volume and tetragonality, due to the small V size.
\\
\textit{Summary} The introduction of V in \ce{BaTiO3} reduces the polarization by overall 30~$\%$. The addition of electrons reduces the expected polarization loss compared with the background-charge reference because of charge-carrier localization on the V~site, an effect which has also been reported in the literature~\cite{Chandra2013}. V doping in \ce{BaTiO3} has been observed experimentally to increase the polarization for low V concentrations ($<$ 0.5 at$\%$), followed by a polarization loss at higher concentrations~\cite{Cai2011}, consistent with our calculations.

\subsubsection{Acceptor Doping}

\paragraph{A-Site Dopants}
Now we move to acceptor doping and begin with the substitution of monovalent K on the Ba A~site. In the K-doped \ce{BaTiO3} system (Ba$_{1-x}$K$_x$TiO$_3$, BKTO), the K is present as K$^+$, which is only slightly larger than Ba$^{2+}$.
\\
\textit{Impurity-atom contribution} Adding a K atom with no additional charge carriers causes a polarization gain of 3~$\%$, caused by complex atom rearrangement, which shows the tendency of the system towards tetragonality, as we will further discuss in the \textit{Lattice-constant contribution} section. This results in a displacement of the K$^+$ ion in the $c$ direction, increased displacements of the Ti on the plane below the K -- they have more space to off-center because of the K displacement -- and decreased off-centering of the Ti above the K. 
\\
\textit{Charge-carrier contribution} Electron depletion increases the polarization by a further 2~$\%$. This is in agreement with the effect observed in the background-charge doped \ce{BaTiO3} (2~$\%$). In fact, the introduction of the K atom has no particular influence on the valence band of the system, which has still O 2p character.
\\
\textit{Lattice-constant contribution} The lattice constant relaxation induces a further polarization increase of 19~$\%$. When adding a K atom, the volume of the system stays rather constant, as K$^+$ is only slightly bigger than the Ba$^{2+}$. More interestingly, the tetragonality of the system is considerably increased, explaining the polarization increase. This indicates that structural factors affect the ferroelectricity of the system primarily through their influence on the ferroelectric $c$ axis: the more space in the $c$ direction (rather than the overall volume), the higher the potential for a ferroelectric distortion. 
\\
\textit{Summary} The introduction of K in \ce{BaTiO3} at a concentration 0.125~h/u.c. overall increases the polarization of the system by 24~$\%$, mainly because of the increased tetragonality. We remind the reader that our polarization values are obtained using the Ba Born effective charge of + 2.7 for K and that they are therefore likely to be smaller in practice. 

\paragraph{B-Site Dopants}

Finally we move to the B-site hole dopants and consider two impurities: Al$^{3+}$ which is quite ionic, and Sc$^{3+}$ which is SOJT active.

We begin with the Al-doped \ce{BaTiO3} (BaTi$_{1-x}$Al$_x$O$_3$, BTAO) in which the trivalent Al$^{3+}$ is smaller than Ti$^{4+}$; we expect it therefore to have a strong structural influence.
\\
\textit{Impurity-atom contribution} Introducing an Al atom in \ce{BaTiO3} without its corresponding charge carriers reduces the polarization by 17~$\%$. The Al atom has in fact a reduced off-centering, since it has no d orbitals that would favour its hybridization with the axial oxygen atom. It is nevertheless still slightly displaced compared to its centrosymmetric position because of its small size. The tolerance factor of pure \ce{BaAlO3}, 1.11, indicates that the Al is small enough to rattle. The decreased Al off-centering also reduces the off-centering of the its axial neighboring Ti along the polar axis.
\\
\textit{Charge-carrier contribution} Electron depletion increases the polarization by 2~$\%$, consistent with the background-charge doped reference and the minimal influence of the Al on the electronic structure in the region of the Fermi energy. While the electron depletion on the oxygen atoms around the Al has a small effect, the depleted equatorial oxygens around the Ti ions cause an increased off-centering of the Ti as in the background-charge case. The Ti axial to the Al are particularly affected, because of the high contribution of the oxygen atoms around them to the top of the valence band, and contribute most to the polarization gain.
\\
\textit{Lattice-constant contribution} Upon relaxing the lattice constants, a polarization loss of 21~$\%$ occurs. This is due to a decrease of the volume and tetragonality of the system, reducing the B-site off-centering homogeneously.
\\
\textit{Summary} The introduction of Al into \ce{BaTiO3} reduces the polarization, overall by 36~$\%$. This is mainly due to the non-off-centering of the Al, which has no d orbitals and is therefore not SOJT active, as well as the loss of space through volume contraction and tetragonality decrease.

The last system that we analyze is Sc-doped tetragonal \ce{BaTiO3} (BaTi$_{1-x}$Sc$_x$O$_3$, BTSO), where Sc is present as Sc$^{3+}$. Sc$^{3+}$ is larger than Ti$^{4+}$ so that we expect it to influence the structure of the system.
\\
\textit{Impurity-atom contribution} Adding a Sc without its charge carriers reduces the polarization by 23~$\%$. The transition metal sites causing the ferroelectricity loss are the Sc~site and its axial neighboring Ti along the polar axis. The origin, as in the BTNO system, is the large Sc$^{3+}$ size: The \ce{BaScO3} tolerance factor is 0.99, meaning that, even though it is SOJT active, the Sc is too big to actually displace in the $c$ direction. The Ti axial to the Sc also off-center less. This is due to the Sc influencing its axial environment; as it is almost centrosymmetric, there is less space for the atoms around it to off-center.
\\
\textit{Charge-carrier contribution} Electron depletion increases the polarization by less than 1~$\%$, a smaller amount compared to the background-charge doped reference (2~$\%$). The oxygen atoms contributing most to the top of the valence band, and therefore most depleted upon doping, are the equatorial and axial oxygens around the Sc. However, because of its large size, the Sc is not affected by this depletion, reducing the net polarization gain.
\\
\textit{Lattice-constant contribution} When the lattice constants of the system are relaxed, the polarization increases by 4~$\%$. This is related to the volume expansion due to the Sc$^{3+}$ being bigger than the Ti$^{4+}$. The tetragonality of the system stays constant and therefore neither counterbalances the volume effect nor increases it.
\\
\textit{Summary} The net polarization loss in BTSO amounts to 19~$\%$. This is due to the large size of the Sc counteracting the other factors that would increase the ferroelectricity of the system.

\subsubsection{Overview Impurity Doping}
Our results suggest that the factors contributing to the polarization change can be divided into three effects: i) the chemistry and size of the impurity atom, ii) the charge carriers and their degree of localization, and iii) changes in tetragonality and volume.

The influence of the impurity atom depends on its size and SOJT activity. On the A~site, small dopants tend to increase the polarization by cation displacement (e.g.~La). Note that the polarization increase due to doping with small A-site cations cannot be achieved if all the A~sites are substituted, where cooperative rotations would be preferred over the ferroelectric distortion (as seen in e.g.~\ce{SrTiO3})~\cite{Collignon2019a}. On the B~site, off-centering is controlled by the presence of empty d orbitals (for the atom to be SOJT active) and the space available for off-centering, which correlates with the size of the atom. The off-centering of SOJT active atoms can be suppressed if they are too large (e.g. Nb and Sc) and does not occur if they lack empty d orbitals in an appropriate energy range (e.g. Al). If both SOJT activity and small size are present, increased off-centering is observed (e.g. V). A theoretically promising (although experimentally inaccessible) avenue would correspond to doping \ce{BaTiO3} with a SOJT active acceptor atom that is smaller than Sc. This is however not feasible as Sc$^{3+}$ is the smallest trivalent SOJT active ion that exists.

The effect of the charge carriers depends on their type and degree of localization. Electrons have a stronger influence on the polarization than holes, as already discussed for the background-charge doped systems. Overall, electrons reduce the polarization, while charge carrier localization can increase (e.g. Nb) or decrease (e.g. V) the extent of the reduction. In contrast, the introduction of holes very slightly increases the polarization of the system at 0.125~carriers/u.c.

The relaxation of the lattice constants permits changes in the tetragonality and volume of the system depending on the introduced dopant. Small atoms tend to induce a volume contraction, reducing the polarization; conversely, large atoms expand the volume and increase the polarization. The determining factor for the polarization change is the gain or loss of space in the ferroelectric direction, which results in a change in tetragonality through the coupling to strain.

We find that the contributions from the impurity atom, the charge carriers and the free lattice constants are additive. At 0.125~carriers/u.c., the polarization change ranges from a gain of 25~$\%$ to a loss of 100~$\%$ (compared to pure \ce{BaTiO3}), depending on the introduced impurity. Which of the contributions dominates depends on the dopant. A general trend can nevertheless be recognized from Figure \ref{fig:polarization_change_percentage}. For donor dopants, the charge carriers are a dominating factor for the ferroelectricity loss, in contrast to acceptor doping, where the chemistry and size of the impurity atom have a prevailing contribution. In both doping regimes, the changes in tetragonality and volume have a strong influence on the system behavior.

Based on our calculations, we can classify the investigated impurity-doped systems with 0.125 carriers/u.c. into one of three categories, based on structural and electronic considerations. In the electron-doped systems, BLTO can be considered to be metallic and non-polar. In fact, it has homogeneously spread out electron charge carriers and a centrosymmetric crystal structure, since the ferroelectric distortion is completely suppressed. BTNO and BTVO can be regarded as being polar-metal-like with a polaronic conduction mechanism, due to the electron localization on the impurity atom or its axial neighboring Ti along the polar axis. They show both a ferroelectric structural distortion, have a non-zero density of states at the Fermi level and the polaronic electron localization could enable polarization switching. The hole-doped systems can also be considered as being polar-metal-like, and in the case of K doping their polarization is even increased over the undoped case.

\section{Conclusions}
In conclusion, we demonstrated computationally that \ce{BaTiO3} can sustain the combination of ferroelectric and metallic properties. With the background-charge method, we found that at low carrier concentrations ($<$~0.2 carriers/u.c.), hole doping increases the polarization, whereas electron doping reduces it. At higher concentrations, both electrons and holes reduce the ferroelectricity of the system by suppressing the B-site off-centering. The effect of electrons is stronger than that of holes because of the nature of the conduction and valence bands: The electrons strongly affect the Ti ions, whereas the holes spread out over both equatorial and axial oxygens. We found that in impurity-doped systems, multiple structural and electronic factors contribute to the polarization change. The contributions can be separated into effects coming from the chemistry and size of the impurity atom, from the charge carriers and from changes in the shape and volume of the system. These contributions are additive and can overall yield a gain (K) or loss (La, Nb, V, Al, Sc) of polarization, depending on the introduced dopant. Based on our calculations, we propose a classification of the investigated materials at low doping concentration as either non-polar metallic (La) or polar-metal-like (Nb, V, K, Al, Sc).

Based on our findings, we propose following experiments to further investigate the interplay between ferroelectricity and metallicity in \ce{BaTiO3}. Field effect doping could contribute to understanding the effect of charge carriers in \ce{BaTiO3}, as predicted by our calculations with background charges. Furthermore, doping of \ce{BaTiO3} with Nb or V could be particularly promising for polarization switchability, which could be allowed by the polaronic localization of the charge carriers in these systems.

\section*{Acknowledgments}
This work was funded by the European Research Council under the European Union’s Horizon~2020 research and innovation program Grant Agreement No.~810451, the Körber Foundation and ETH Zürich. V. F. M. was supported by a MARVEL INSPIRE Potentials Master's Fellowship from the NCCR MARVEL, funded by the Swiss National Science Foundation. Calculations were performed on the ETH Zürich Euler cluster

\bibliography{references}

\onecolumngrid
\newpage
\appendix*
\section{Supplementary Information}
If not otherwise specified in the caption of the figures, the computational parameters used correspond to the ones described in the methods section of the main text.

\subsection{Undoped Tetragonal \ce{BaTiO3}}
\subsubsection{Berry-phase Calculation}
\begin{figure}[h]
    \centering
    \includegraphics[scale=1]{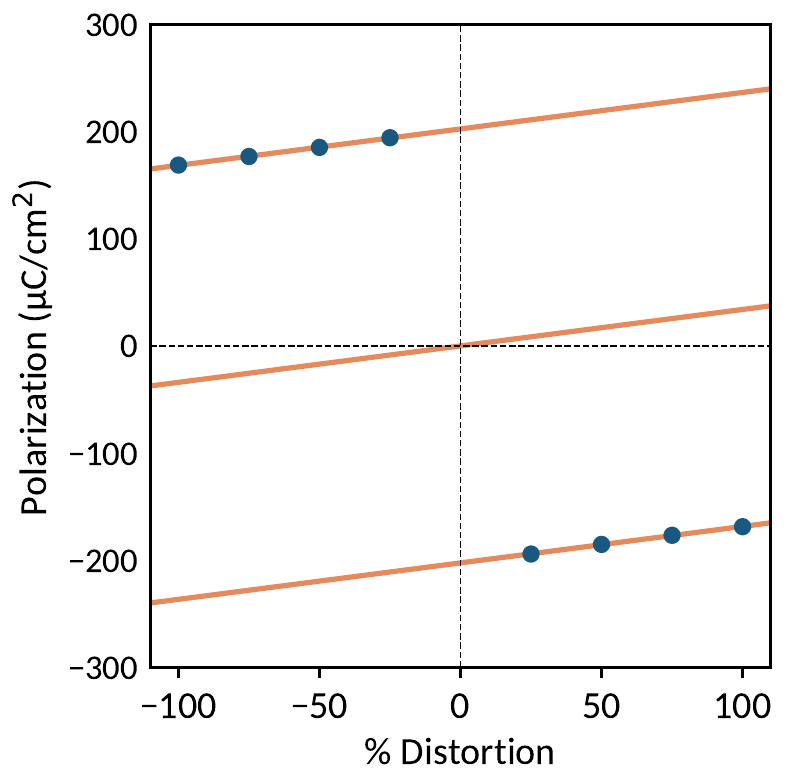}
    \caption{Berry-phase calculation for tetragonal \ce{BaTiO3}. The calculated spontaneous polarization of the system is 34~$\upmu$C/cm$^2$.}
    \label{fig:berry_phase}
\end{figure}

\newpage
\subsection{Background-charge Doped Pseudocubic \ce{BaTiO3} Unit Cell}
\subsubsection{Density of States}
\begin{figure}[h]
    \centering
    \includegraphics[scale=1]{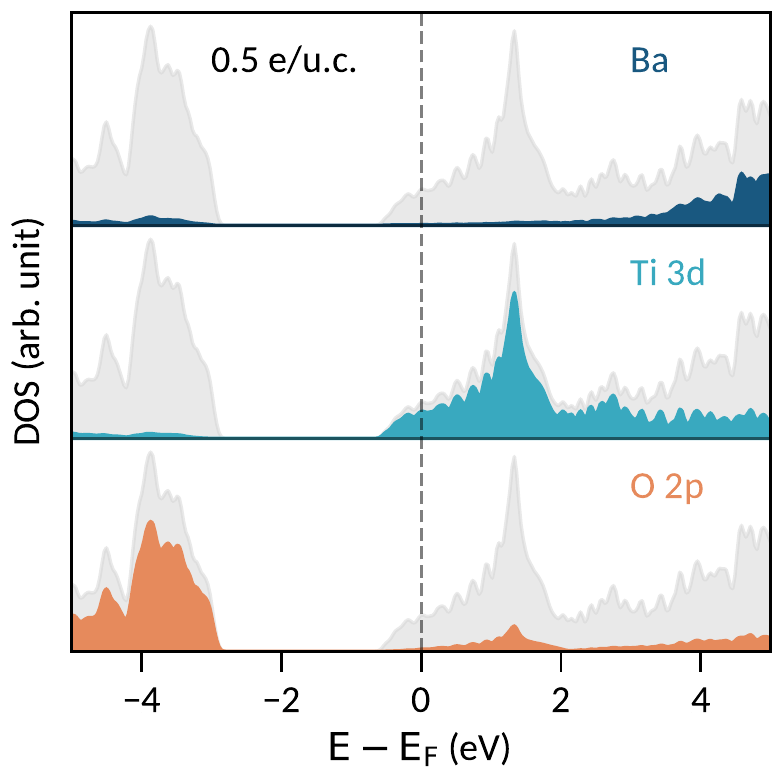}
    \caption{Density of states for pseudocubic \ce{BaTiO3} doped with 0.5~e/u.c.}
    \label{fig:dos_cub_bg_0.5e}
\end{figure}
\begin{figure}[h]
    \centering
    \includegraphics[scale=1]{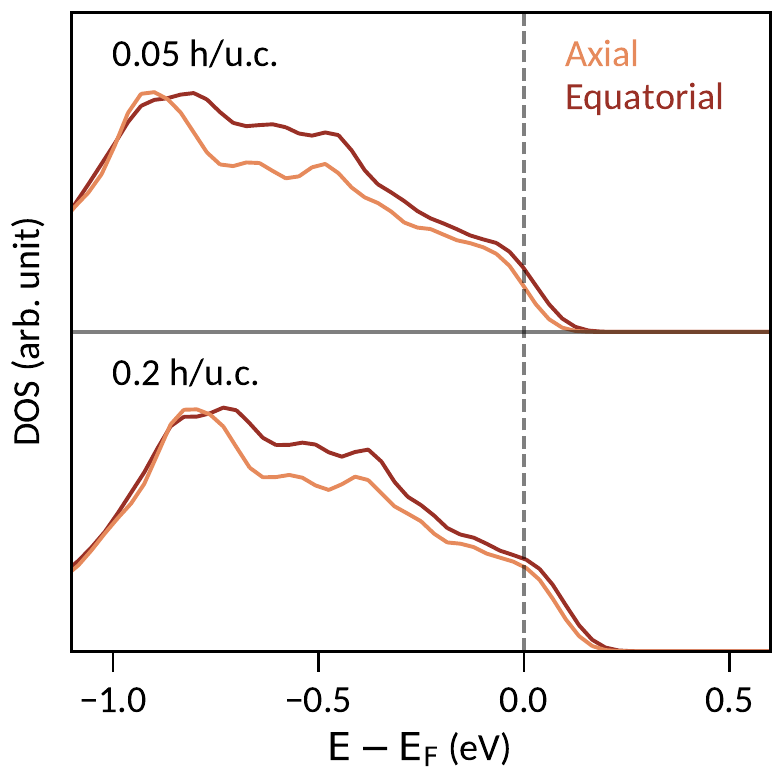}
    \caption{Oxygen site-resolved densities of states for pseudocubic \ce{BaTiO3} doped with 0.05~h/u.c. (top) and 0.2~h/u.c. (bottom).}
    \label{fig:dos_cub_bg_h}
\end{figure}

\newpage
\subsection{Impurity-doped \ce{BaTiO3} Supercells}
\subsubsection{Polarization, Volume and Tetragonality}
\begin{figure*}[hp]
    \centering
    \includegraphics[scale=0.8, keepaspectratio]{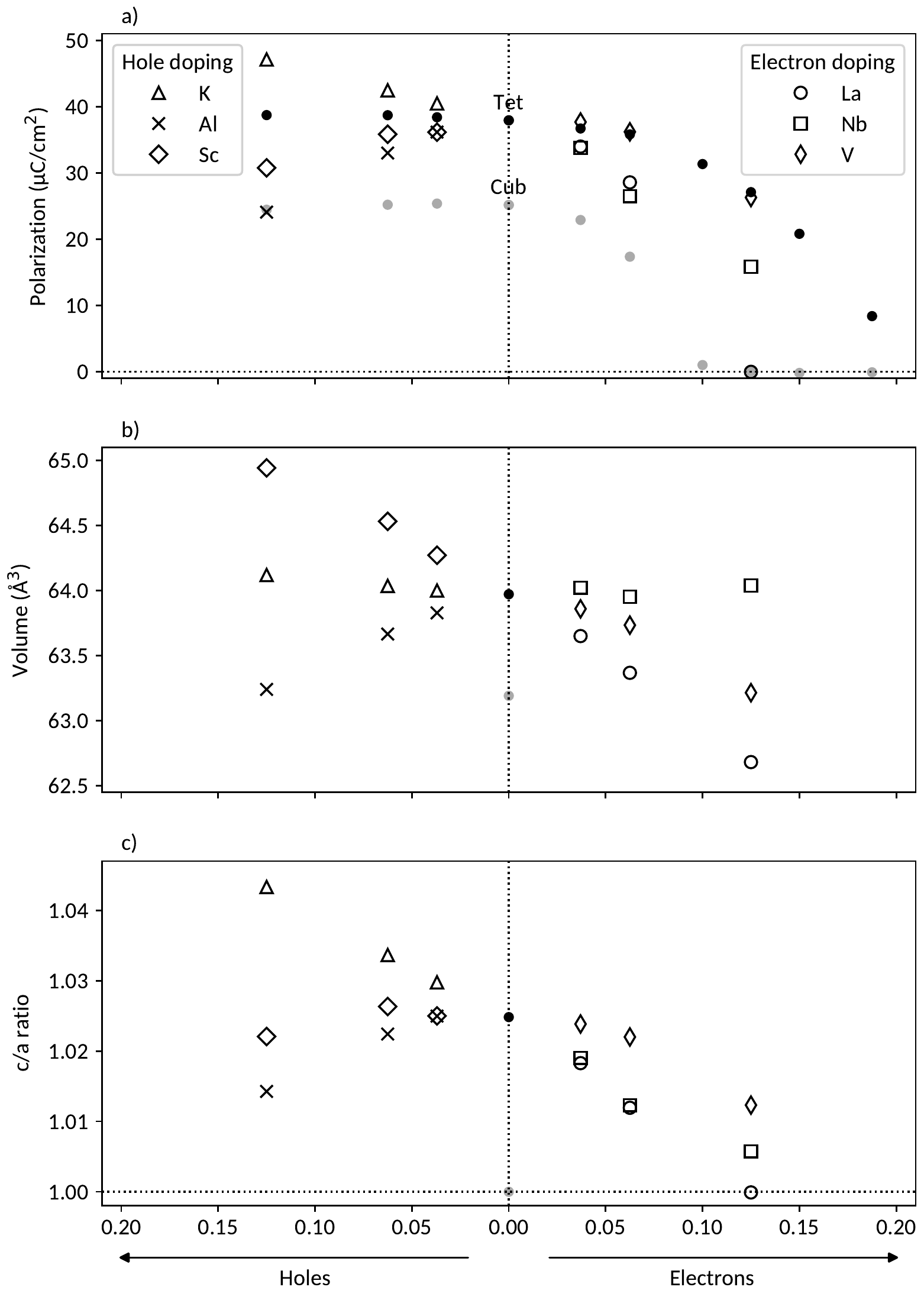}
    \caption{Polarization (a), average volume (b) and $c/a$ ratio (c) of doped \ce{BaTiO3} as a function of the charge-carrier concentration (electrons on the right, holes on the left). The charge carriers are introduced through doping with impurity atoms in 2$\times$2$\times$2, 2$\sqrt{2}$$\times$2$\sqrt{2}$$\times$2 and 3$\times$3$\times$3 supercells and their charge-carrier concentrations are 0.125, and 0.0625 and 0.037~carriers/u.c., respectively. The k-point grids used for these calculations are 12$\times$12$\times$12, 6$\times$6$\times$6 and 4$\times$4$\times$4, respectively. The systems are labeled with the name of their respective dopant. The black and gray dots correspond to the tetragonal and pseudocubic background-charge references.}
    \label{fig:pol_vol_c_a_supercells}
\end{figure*}

\newpage
\subsubsection{Bond Ratios}
\begin{figure*}[hp]
    \centering
    \includegraphics[scale=0.8, keepaspectratio]{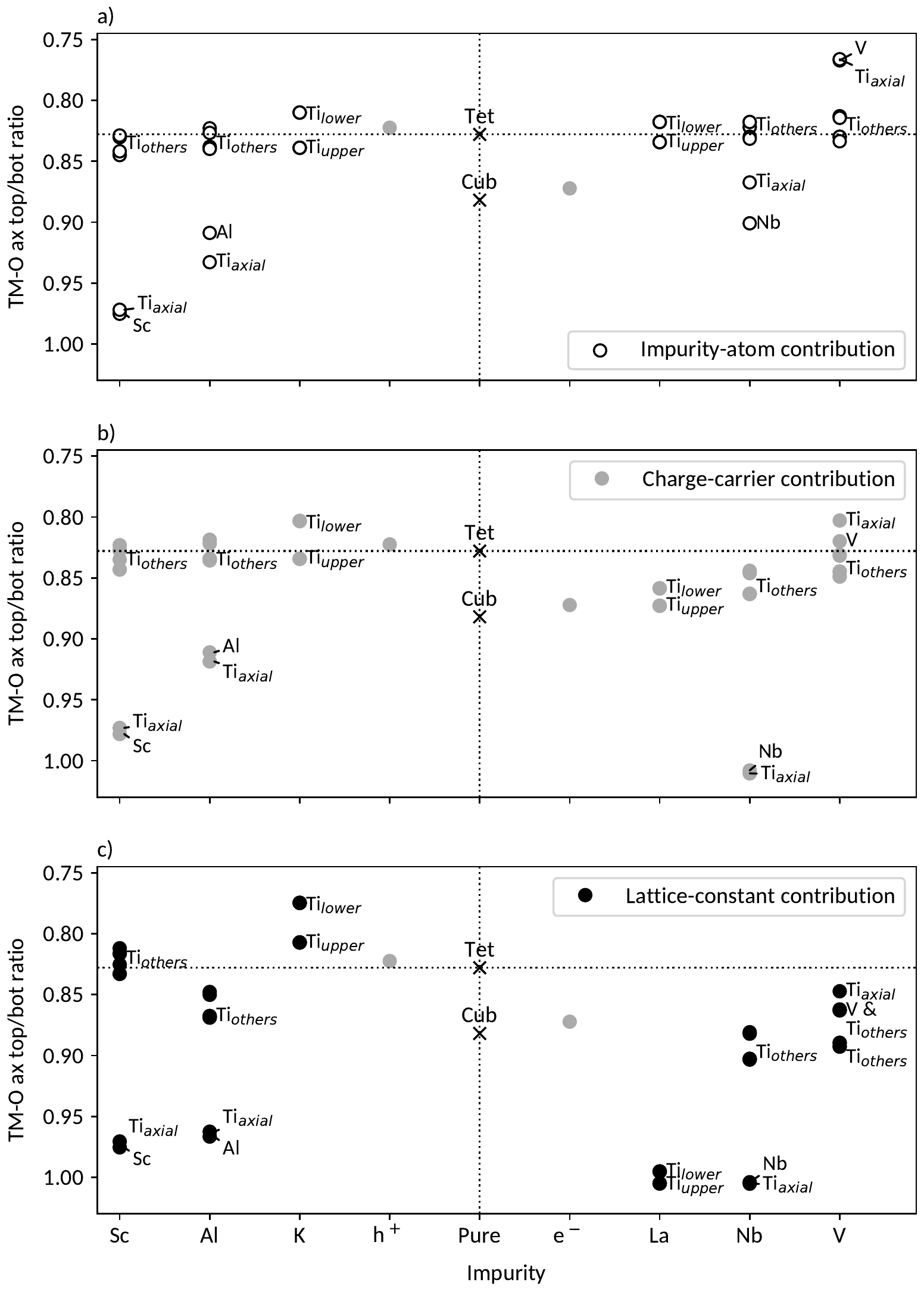}
    \caption{Bond ratios of the transition metal - oxygen axial top to bottom bonds for tetragonal \ce{BaTiO3}. a) Impurity-atom contribution with the impurity atom, 0~carriers/u.c. and fixed lattice constants (white circles), b) Charge-carrier contribution with the impurity atom, 0.125~carriers/u.c. and fixed lattice constants (gray circles) and c) Lattice-constant contribution with the impurity atom, 0.125~carriers/u.c. and free lattice constants (black circles). The impurity atom is indicated on the x-axis. Each system corresponds to a 2$\times$2$\times$2 supercell and therefore has eight bond ratio values (for its eight transition metals), partly overlapping. Pure \ce{BaTiO3} in tetragonal and pseudocubic symmetry (abbreviated \textit{tet} and \textit{cub}) as well as background-charge doped tetragonal systems (holes: h$^+$, electrons: e$^-$) are given as a reference.}
    \label{fig:bond_ratios_supercells}
\end{figure*}

\newpage
\subsubsection{Density of States}
\begin{figure*}[hp]
    \centering
    \includegraphics[width=17.8cm, keepaspectratio]{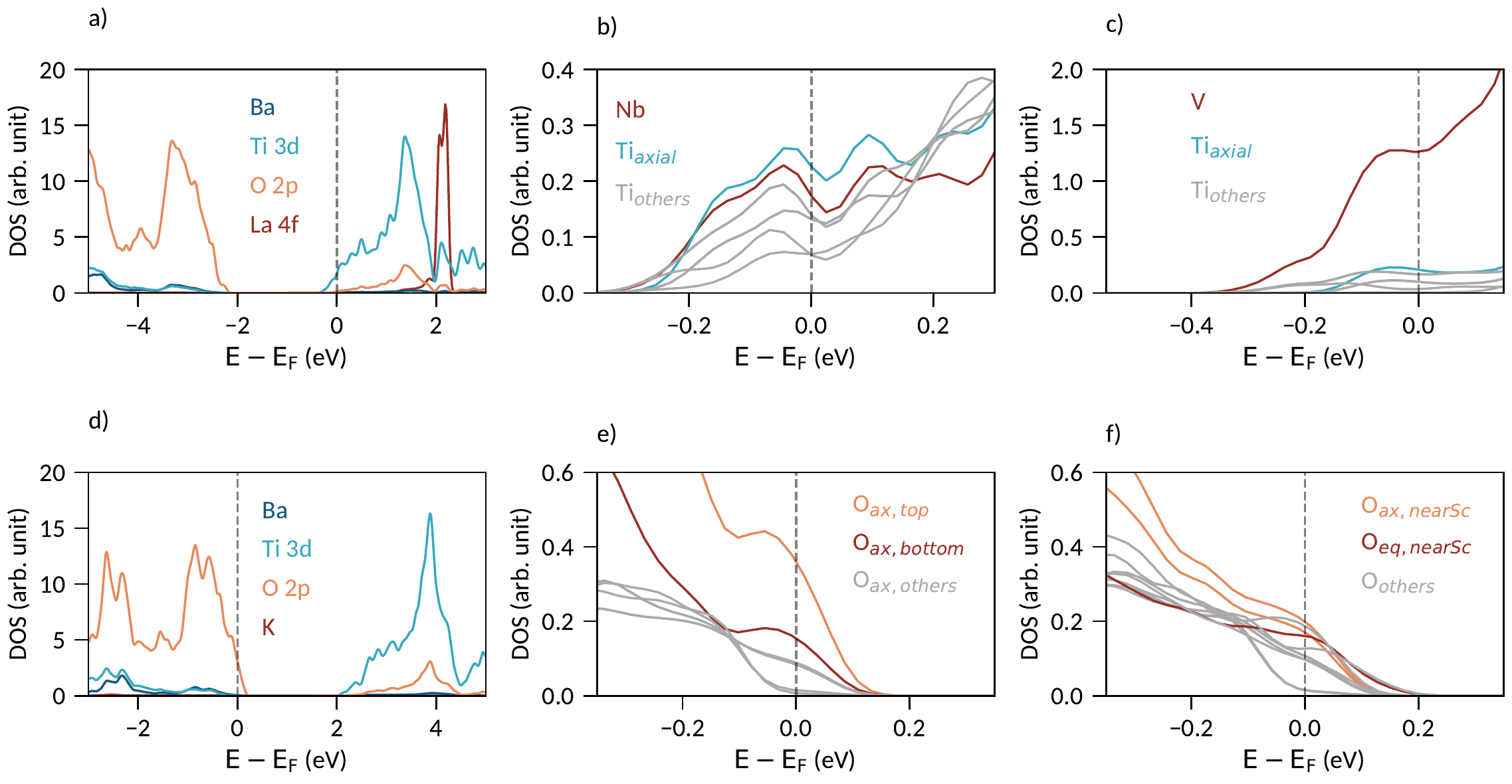}
    \caption{a) Density of states for La-doped \ce{BaTiO3} (with one La in a 2$\times$2$\times$2 \ce{BaTiO3} supercell with fixed lattice constants). As the La introduces one valence electron into the system, the Fermi energy lies in the conduction band. The La f-states lie at higher energies and the main contribution to the conduction band comes from the Ti d orbitals. b) Transition-metal site-resolved density of states of Nb-doped \ce{BaTiO3} (BTNO with one Nb in a 2$\times$2$\times$2 \ce{BaTiO3} supercell and fixed lattice constants). The region around the Fermi energy in the conduction band is shown. Two sites, the Nb and its axial neighboring Ti along the polar axis, have particularly large contributions. c) Transition-metal site-resolved density of states for V-doped \ce{BaTiO3} (with one V in a 2$\times$2$\times$2 \ce{BaTiO3} supercell with fixed lattice constants). The conduction band region around the Fermi energy is shown. The V has a particularly large contribution, whereas the its axial neighboring Ti along the polar axis hardly contributes to the conduction band. d) Density of states for K-doped \ce{BaTiO3} (BKTO with one K in a 2$\times$2$\times$2 \ce{BaTiO3} supercell and fixed lattice constants). One valence electron is depleted from the system so that the Fermi energy lies in the valence band. The K does not affect the DOS around the Fermi energy. e) Axial oxygen site-resolved density of states of Al-doped \ce{BaTiO3} (with one Al in a 2$\times$2$\times$2 \ce{BaTiO3} supercell with fixed lattice constants). The most contributing sites are the axial oxygens around the Al atom (top and bottom). f) Oxygen site-resolved density of states of Sc-doped \ce{BaTiO3} (BTSO with one Sc in a 2$\times$2$\times$2 \ce{BaTiO3} supercell and fixed lattice constants). The most contributing sites are the axial and equatorial oxygens around the Sc atom. All the supercells have tetragonal symmetry. All DOS are computed with a 12$\times$12$\times$12 k-point grid.}
    \label{fig:dos_supercells}
\end{figure*}

\end{document}